%% file: main.tex
\documentclass{article}             
\input{preamble}

\title{Going in Style: Audio Backdoors through Stylistic Transformations}
 
\name{Stefanos Koffas$^{1,*}$, Luca Pajola$^{2,*}$, Stjepan Picek$^{3,1}$, Mauro Conti$^{2,1}$}    
 \address{$^1$Cybersecurity Group, Delft University of Technology, The Netherlands\\
          $^2$Security and Privacy Research Group, University of Padua, Italy\\
          $^3$Digital Security Group, Radboud University, The Netherlands
          }
\begin{document}
\maketitle

\begin{abstract}
This work explores \textit{stylistic} triggers for backdoor attacks in the audio domain: dynamic transformations of malicious samples through guitar effects. We first formalize stylistic triggers -- currently missing in the literature. 
Second, we explore how to develop stylistic triggers in the audio domain by proposing \textit{JingleBack}. Our experiments confirm the effectiveness of the attack, achieving a 96\% attack success rate. Our code is available in \url{https://github.com/skoffas/going-in-style}. 
\end{abstract}

\input{Sections/01-Introduction}
\input{Sections/02-Background}
\input{Sections/03-Threat-Model}

\input{Sections/06-Case-study-speech-classification}
\input{Sections/08-Conclusions}

\bibliographystyle{ieeetr}
\bibliography{sample}

\end{document}

%% file: preamble.tex
\usepackage{icassp}
\usepackage[switch]{lineno}

\usepackage[english]{babel}

\usepackage{subcaption}
\usepackage{amsmath}
\usepackage{graphicx}
\PassOptionsToPackage{hyphens}{url}\usepackage[hidelinks]{hyperref}
\usepackage[noabbrev,nameinlink]{cleveref}

\usepackage{array}
\usepackage{soul,xcolor, colortbl}
\usepackage{multirow}
\usepackage{tabularx}
\usepackage{paralist}
\newcolumntype{P}[1]{>{\centering\arraybackslash}p{#1}}

\usepackage{todonotes}

\usepackage{booktabs}
\usepackage{tikz}

\newcommand\blfootnote[1]{%
  \begingroup
  \renewcommand\thefootnote{}\footnote{#1}%
  \addtocounter{footnote}{-1}%
  \endgroup
}

%% file: Sections/01-Introduction.tex
\section{Introduction}
\label{sec:introduction}
\blfootnote{$^*$Equal Contribution}
Backdoor attacks are a class of machine learning threats where the attacker embeds a secret functionality into the victim's model, which can be triggered at the testing time from malicious inputs~\cite{gu2017badnets, zhai2021backdoor}.
Backdoor triggers can be grouped into two major families: \textit{static}, when the trigger is a fixed pattern attached to the poisoned sample~\cite{gu2017badnets}, and \textit{dynamic} when the trigger's properties vary for each poisoned sample~\cite{dynamic-backdoor-attacks-against-ml-models,dynamic-backdoors-with-gap}. Dynamic triggers are generally stronger as they can be effective under different conditions and potentially bypass state-of-the-art countermeasures~\cite{dynamic-backdoor-attacks-against-ml-models}.
Recently, a new way of creating dynamic triggers has emerged: stylistic triggers.
For instance, paintings and writing styles can be used as triggers in computer vision (CV)~\cite{deep-feature-space-trojan-attack-of-nns-by-controlled-detoxification} and text domain~\cite{mind-the-style-of-text}.

\textbf{Motivation.}
Stylistic backdoors are powerful but not yet studied in the audio domain. Such backdoors highly depend on the targeted application resulting in domain-specific challenges. In audio, the most important challenge is to create a style that alters the original signal in a way that is distinguishable by the trained models but also keeps the audio quality at an acceptable level. To our best knowledge, no work has explored stylistic backdoors in audio, and we aim to fill this gap.
We investigate the effect of six stylistic triggers in a speech classification task in both clean and dirty-label settings.
Triggers are generated through electric guitar effects that do not require training of complex generative models~\cite{deep-feature-space-trojan-attack-of-nns-by-controlled-detoxification,mind-the-style-of-text}, making our attack easy to deploy.
Our contributions are:
\begin{compactitem}
    \item We propose and demonstrate the feasibility of stylistic backdoor attacks (denoted JingleBack) in the audio domain through electric guitar effects. For our experiments, we trained 444 models, and JingleBack reached up to 96\% attack success rate.
    \item We are the first to formally describe stylistic backdoor attacks and establish a domain-agnostic framework that can be used for stylistic backdoors in any application.
\end{compactitem}

%% file: Sections/02-Background.tex
\section{Background}
\label{sec:background}

\textbf{Evasion Attack.}
The attacker aims to find a small perturbation $\epsilon$ for a sample $x$ to produce a misclassification in a target classifier $\mathbf{C}$.
The definition of $\epsilon$ is task-dependent~\cite{evasion2004}.
For images, $\epsilon$ is usually an additive noise at a pixel level computed via a gradient-based procedure~\cite{fgsm}.
For audio, $\epsilon$ is a small waveform computed through an optimization process~\cite{audio-adversarial-examples-targeted-attacks-on-speech-to-text}.

\textbf{Poisoning Attack.}
Attackers able to manipulate victims' training data $\mathcal{D}^{tr}$ can inject adversarial samples to increase $\mathbf{C}$'s testing error and decrease the model's performance~\cite{biggio_poisoning}.
Data manipulation is a concrete threat because building datasets often involves untrusted sources, and complex model training is outsourced to third parties.

\textbf{Backdoor Attack.}
Data poisoning leads to \textit{backdooor attacks}. The attackers insert backdoors in the model resulting in targeted misclassifications~\cite{gu2017badnets}.
Most backdoors are source-agnostic, as the trigger can be applied to any dataset's class. There are two different approaches to creating such a backdoor: \textit{dirty-label}~\cite{gu2017badnets}, where the adversary adds the trigger to some training samples but also alters their labels, and \textit{clean-label}~\cite{clean-label-backdoor-attacks}, where the attacker only poisons samples from the target class.
The dirty-label attack produces stronger backdoors and requires less poisoned data, but the clean-label attack is a more realistic threat as poisoned samples cannot be easily identified by manual inspection~\cite{clean-label-backdoor-attacks}.
For example, crowdsourced datasets like Mozilla's Common Voice~\cite{common-voice-dataset} use volunteers to validate each audio file and its transcription.
When the attack is dirty-label, the volunteer can easily spot inconsistencies between the provided audio and its transcription. However, in a clean-label attack, the trigger will remain unnoticed because the audio and its transcription will be correct.
This work explores the effects of both approaches on a style-based backdoor attack in speech classification.

%% file: Sections/03-Threat-Model.tex
\section{Style Backdoor Attack}
\label{sec:threat}

\subsection{Threat Model}

\textbf{Attacker's Capabilities.}
We assume a gray-box threat model where the adversary has access to a small portion of the training data, which can be altered without restriction, but has no knowledge of the training algorithm and the model's architecture. 
Such capabilities are realistic as modern datasets are based on crowdsourcing~\cite{common-voice-dataset}, and malicious data may evade security checks~\cite{large-dataset-pyrrhic-win}. 

\textbf{Attacker's Aim.}
The attacker's purpose is to insert a secret functionality into the deployed model, which is activated when the trigger is present in the model's input. Usually, this functionality causes targeted misclassifications with a very high probability. Furthermore, the model should behave normally for any other input to avoid raising suspicions. 

\subsection{Attack Formulation}
Prior studies in backdoor attacks mainly focus on \textit{static triggers}, where the attack trivially inserts a trigger $\epsilon$ in victim samples.
Such addition is sample-independent.
Let $x$ be a sample, $\epsilon$ a trigger, the backdoor $\mathcal{F}$ can be described as: 
\begin{equation}\label{eq.backdoor_static}
\setlength\abovedisplayskip{0pt}
\setlength\belowdisplayskip{0pt}
    \mathcal{F}(x, \epsilon) = x + \epsilon.  
\end{equation}
In BadNets~\cite{gu2017badnets} (CV), the trigger is a patch at a pixel level that replaces a part of the original sample. In audio, a static trigger could be a tone superimposed on the original sample~\cite{can-you-hear-it}.
Static triggers can thus be considered as a \textit{constant} parameter of the attack, as a batch of benign samples $\{x_0, \ldots, x_n\}$ is transformed in nothing more than $\{x_0 + \epsilon, \ldots, x_n + \epsilon\}$. 
Thus, the victim's model $\mathbf{C}_{back}$ learns to associate the static pattern $\epsilon$ to a target label $y*$.

In this work, we approach the backdoor attack from an orthogonal perspective: can the trigger be something that the sample \textit{is} rather than something the sample \textit{has}? 
Static backdoors answer the \textit{has} proposition: the poisoned sample contains a specific (and constant) pattern that $\mathbf{C}_{back}$ associates with the target class.
Answering the \textit{is} is more complex: this is how a sample is presented, and we can think of it as a latent variable globally describing that sample. 
In our study, global descriptors are defined by \textit{stylistic properties}.
Examples are image exposure (CV), writing complexity (text), and the signal's pitch (audio).
Once the stylistic trigger $\epsilon$ is identified, we need a function $\mathcal{S}_\epsilon$ that embeds such a style to a given sample $x$. Formally, Eq.~\eqref{eq.backdoor_static} changes to:
\begin{equation}\label{eq.backdoor_style}
\setlength\abovedisplayskip{0pt}
\setlength\belowdisplayskip{0pt}
    \mathcal{F}(x, \mathcal{S}_\epsilon) = \mathcal{S}_\epsilon(x).  
\end{equation}
We explore different $\mathcal{S}_\epsilon$ for speech recognition in the following sections. 
With the formulation given in Eq.~\eqref{eq.backdoor_style}, the batch of adversarial samples$\{\mathcal{S}_\epsilon (x_0), \ldots,$ $\mathcal{S}_\epsilon (x_n)\}$ now contains dynamic triggers since the outcome of the backdoor generation varies based on the inputted sample. 
Now, suppose that $\mathcal{S}_\epsilon$ exists.
The stylistic backdoor is effective if the model $\mathbf{C}_{back}$ learns an association between $\mathcal{S}_\epsilon$ and the target class $y*$.
This problem can be formulated with two sub-questions:
\begin{compactenum}
    \item Can $\mathbf{C}$ learn to recognize the presence of $\mathcal{S}_\epsilon$? We need proof or at least an indication that stylistic properties can be learned by the victim's model.  
    \item Can we let $\mathbf{C}$ learn the association between $\mathcal{S}_\epsilon$ and $y*$? If the previous question is positively answered, we need to understand further how to create the backdoor and what are the possible obstacles at this stage.  
\end{compactenum}
Geirhos et al.~\cite{geirhos2018imagenettrained} answered the first question, showing that CNNs focus more on textures rather than object shapes.
For example, an image with `cat' shapes and `elephant' texture is classified as an elephant. 
Similarly, styles can be learned in the speech domain~\cite{grinstein2018audio, AlBadawy2020}.

Finally, we need to understand what are the conditions to create the stylistic backdoor in $\mathbf{C}$.
In source-agnostic backdoors, the trigger is present only in one of the classes of the training data. 
This condition is easily satisfied by the classic backdoor attacks injecting artifacts as triggers.
Conversely, with stylistic backdoors, such conditions might not always be met. 
For example, high exposure (image domain) and low-frequency tunes (audio domain) might be present in many classes of clean data. 
We conclude that the choice of $\mathcal{S}_\epsilon$ impacts attack success. 
In our experiments, we use stylistic functions that are unlikely to be present in the training data. 

%% file: Sections/06-Case-study-speech-classification.tex
\section{JingleBack Design}
\label{sec:speech}

\subsection{Stylistic Generation}
\label{ssec:sound-style-gen}
This work focuses on backdoors in audio aiming into an easy-to-deploy attack. Thus, we investigate if such an attack can be implemented through simple digital music effects like chorus or gain.
We used Spotify's pedalboard\footnote{\url{https://github.com/spotify/pedalboard}} to implement six styles by combining effects like PitchShift, Distortion, Chorus, Reverb, Gain, Ladderfilter, and Phaser. We explain the adopted effects briefly and show their analytical mathematical expressions in~\Cref{tab:equations}.

\textbf{PitchShift} increases the pitch of the original audio by a number of semitones without affecting its duration. The pitch is the lowest frequency $f_0$ of a signal $S$~\cite[p.~xv]{music-in-theory-and-practice}. A semitone $sem$ is the smallest musical interval used in music denoting a different tone.
\textbf{Distortion} applies a distortion with a $tanh()$ waveshaper. $\beta$ controls the signal's amplitude increase.
\textbf{Chorus} imitates a group of musicians that play the same sound
by superimposing many versions of the same sound that are slightly out of time and tune. Pedalboard's chorus uses one unsynchronized version with the provided delay $d$ where $\alpha$ controls the chorus's amplitude.
\textbf{Reverberation} imitates the reflection of reproduced sound on various surfaces.
Spotify's pedalboard is based on FreeVerb\footnote{\url{https://ccrma.stanford.edu/~jos/pasp/Freeverb.html}}, which uses eight parallel Schroeder-Moorer filtered-feedback comb-filters that create eight delayed versions of the input signal that are added and fed into four Schroeder all-pass filters in series. In~\Cref{tab:equations}, $AP_{1-4}$ is the combined effect of the four cascaded all-pass filters, and $CF_{i}$ is the $i^{th}$ comb-filter.
\textbf{Gain}
changes the signal amplitude by a factor $G$.
Spotify pedalboard implements a Moog \textbf{Ladder} filter $L(\cdot)$~\cite{moog-ladder-filter}. In our experiments, we use the \textit{high-pass} version as it had a clear effect on our samples.
\textbf{Phaser} is based on special filters that can change the frequencies they block over time through a low-frequency oscillator $P$. The phaser superimposes the original signal with its altered version.
This results in a soft-moving sound.
$\alpha$ controls the effect's intensity.

\begin{table}[t]
    \centering
    \caption{Equations of the chosen effects.}
    \vspace{-3mm}
    \label{tab:equations}
    \resizebox{0.25\textwidth}{!}{%
        \begin{tabular}{c}
        \toprule
        \textbf{Effect Equation}\\ \toprule
        $\mathrm{PitchShift}(S(f_0), sem) = S(f_0\cdot e^{sem/12})$ \\ \hline
        $\mathrm{Distortion}(S(t), \beta) = \beta \cdot tanh(S(t))$\\ \hline
        $\mathrm{Chorus}(S(t), d, \alpha) = S(t) + \alpha \cdot S(t - d)$\\ \hline
        $\mathrm{Reverb}(S(t)) = AP_{1-4}[\sum_{i=1}^{8}CF_{i}(S(t))]$\\ \hline
        $\mathrm{Gain}(S(t), G) = G\cdot S(t)$\\ \hline
        $\mathrm{Ladder}(S(t)) = \alpha \cdot L(S(t))$\\ \hline
        $\mathrm{Phaser}(S(t)) = S(t) + \alpha \cdot P(t, S(t))$\\ \bottomrule
        \end{tabular}
    }
    \vspace{-6mm}
\end{table}

\subsection{Experimental Settings}
\label{ssec:sound-setup}

\textbf{Dataset and Features.}
We used Google's Speech Commands (GSC) dataset~\cite{speech-commands-dataset}, containing 30 classes of audio keywords like ``yes'', ``no'', ``up'', and ``down''.
We used the Mel-frequency cepstral coefficients (MFCCs) as input features because they are rather accurate in emulating the human vocal system and widely used~\cite{can-you-hear-it}.
We use common settings described in the literature~\cite{generalized-end-to-end-loss-for-speaker-verification}, i.e., 40-mel bands, a step of 10ms, and a window length of 25ms.

\textbf{Backdoor.}
We split our data into training, validation, and test sets in a 64/16/20 way. We poisoned up to 1\% of the training data and used two backdoor settings: clean-label and dirty-label attacks.
We chose the class ``yes'' as the target class without loss of generality since we expect similar behavior regardless of the target class.
Triggers are generated with the six styles described in~\Cref{tab:styles}.
Parameters are selected to limit sample distortion and preserve their quality.

\begin{table}[h]
    \centering
    \caption{Stylistic triggers deployed in our experiments.}
    \label{tab:styles}
    \vspace{-3mm}
    \resizebox{0.35\textwidth}{!}{%
        \begin{tabular}{c|c}
        \toprule
        \textbf{Style} & \textbf{Effect} \\ \toprule
        0 & $\mathrm{PitchShift}(S, 10)$ \\ \hline
        1 & $\mathrm{Distortion}(S, 30dB)$ \\ \hline
        2 & $\mathrm{Chorus}(S, 10ms, 5)$ \\ \hline
        3 & $\mathrm{Chorus}(\mathrm{Distortion}(\mathrm{PitchShift}(S, 10), 20dB), 8ms, 5)$ \\ \hline
        4 & $\mathrm{Reverb(Distortion(Chorus}(S, 15ms, 0.25), 20dB))$ \\ \hline
        5 & $\mathrm{Phaser(Ladder(Gain}(S, 12dB)))$ \\ \bottomrule
        \end{tabular}
    }
    \vspace{-2mm}
\end{table}

\par
\textbf{User study.}
We conducted a user study (250 samples, 30 participants) to verify that the adversarial samples preserve the original audio.
Each sample is reviewed by three participants.
We analyze the perturbation effect in terms of users' accuracy (random guessing at 3\%). 95\% of legitimate samples are correctly classified, implying that the dataset contains audio that is not always comprehensible by target users. Stylistic transformation, instead, slightly degrades samples quality; in order, from style 0 to style 5, we found the following scores: 71\%, 80\%, 89\%, 45\%, 86\%, and 86\%. 
\par
\textbf{Models.} We used three models, two CNNs (small and large) and one LSTM as described in~\cite{can-you-hear-it}.
Experiments are repeated four times to limit the randomness in results.
Each model is trained for a maximum of 300 epochs, with an early stopping (patience of 20 epochs) based on the validation loss.
In total, by considering stylistic triggers (6), backdoor settings (2), models (3), poisoning rates (3), and repetitions (4), 432 poisoned models are trained. We also trained 12 clean models (4 repetitions for each model) to use as a reference when investigating the backdoor's effect on the original task.

\section{Experimental Results}
\label{ssec:sound-results}

\subsection{Effect on Clean Accuracy}
We first verify that the backdoored models have comparable performance to their clean versions.
Clean models show, on average, high performance (expressed as F1-score): large CNN $93.8\pm0.2$, small CNN $87.2\pm0.3$, and LSTM $90.8\pm1.1$.
We notice that our attack is stealthy since we observed only a small drop in the performance of the backdoored models.
On average, models drop $0.24\pm0.9$ pp (points percentage) in the F1-score, while the worst model drops 4.87 pp.
Among the 432 poisoned models, the performance drop is $>2$ pp in 23 cases and decreases to 10 cases when $>3$ pp.

\subsection{Backdoor}
\begin{figure}[!htpb]
    \centering
    \setlength{\belowcaptionskip}{-20pt}
    \includegraphics[width=0.9\linewidth]{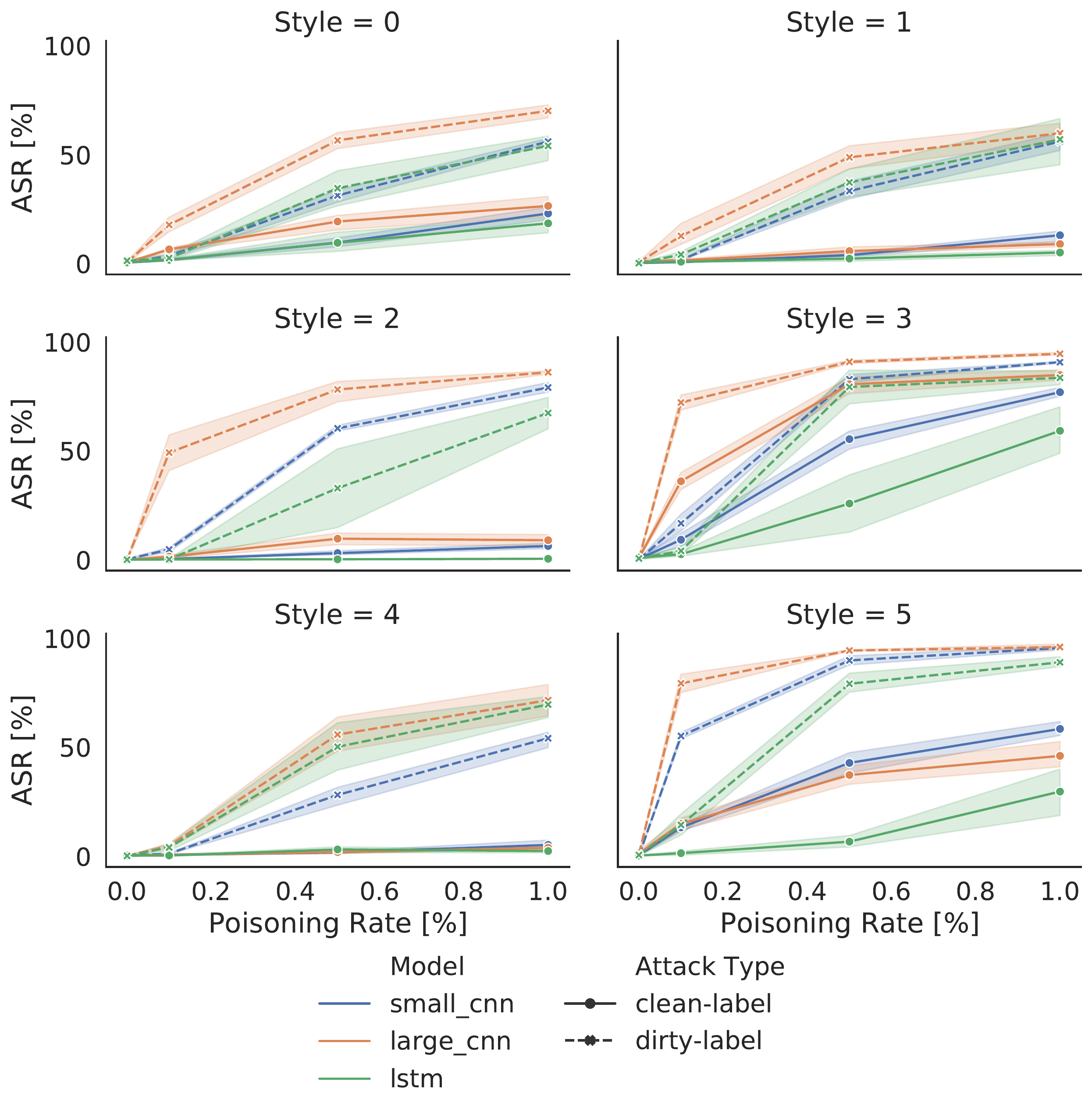}
    \caption{Backdoor attack success rate.}
    \label{fig:asr}
\end{figure}

We analyze the performance of our proposed stylistic backdoor under clean and dirty-label settings. \Cref{fig:asr} shows the results. The y-axis shows the attack success rate (ASR), which represents the percentage of successfully triggered backdoors over a number of tries, and the x-axis shows the poisoning rate, which is the percentage of the poisoned training samples used for our attack.
Our results are in line with the literature~\cite{can-you-hear-it, clean-label-backdoor-attacks}.
For instance, the higher the poisoning rate, the higher the ASR because the poisoned models have more samples to learn the trigger~\cite{can-you-hear-it}.
Additionally, the dirty-label attack is more effective as it almost always results in higher ASR than the clean-label attack, as shown in~\cite{clean-label-backdoor-attacks}.
In particular, dirty-label attacks include cases in which even a small poisoning rate ($0.1\%$) is effective (ASR $>50\%$): examples are style 2 in Large-CNN, style 3 in Large-CNN, and style 5 in both Large-CNN and Small-CNN.
Furthermore, Large-CNN -- the best-performing model on average -- is the most vulnerable, showing that backdoor effectiveness is connected to the model's ability to learn.
\par
Generally, we can notice that styles have different effects (e.g., style 3 is always more effective than style 1), leading to the following observations.
(i) The clean-label attack is effective only with styles 3 and 5.
(ii) Dirty-label attack shows better performance in all cases.
(iii) The addition of effects does not always result in a performance boost (e.g., style 0 outperforms style 4 in clean-label settings).
\par
In the dirty-label attack, the poisoned samples are generally different from the samples of the target class as they originally belonged to different classes. As a result, the distance between these samples in the feature space is large and easy to learn by our models, even by applying simple effects to them. However, in the clean-label attack, the poisoned samples belong to the target class, and there is a higher probability that their features are not very different from the clean samples. For that reason, the dirty-label attack is more effective than the clean-label attack, which explains (i) and (ii). We believe that adding more effects is not the best way to create distinguishable triggers because what is most important is to change the signal's representation in the feature space. For that reason, a simpler transformation that alters the frequency representation of the original signal, like $PitchShift$ may result in a larger difference in the MFCCs as they use Fourier transform internally. This explains (iii).

\subsection{Ablation Study}
\begin{figure}[!htpb]
    \centering
    \setlength{\belowcaptionskip}{-10pt}
    \includegraphics[width=0.36\textwidth]{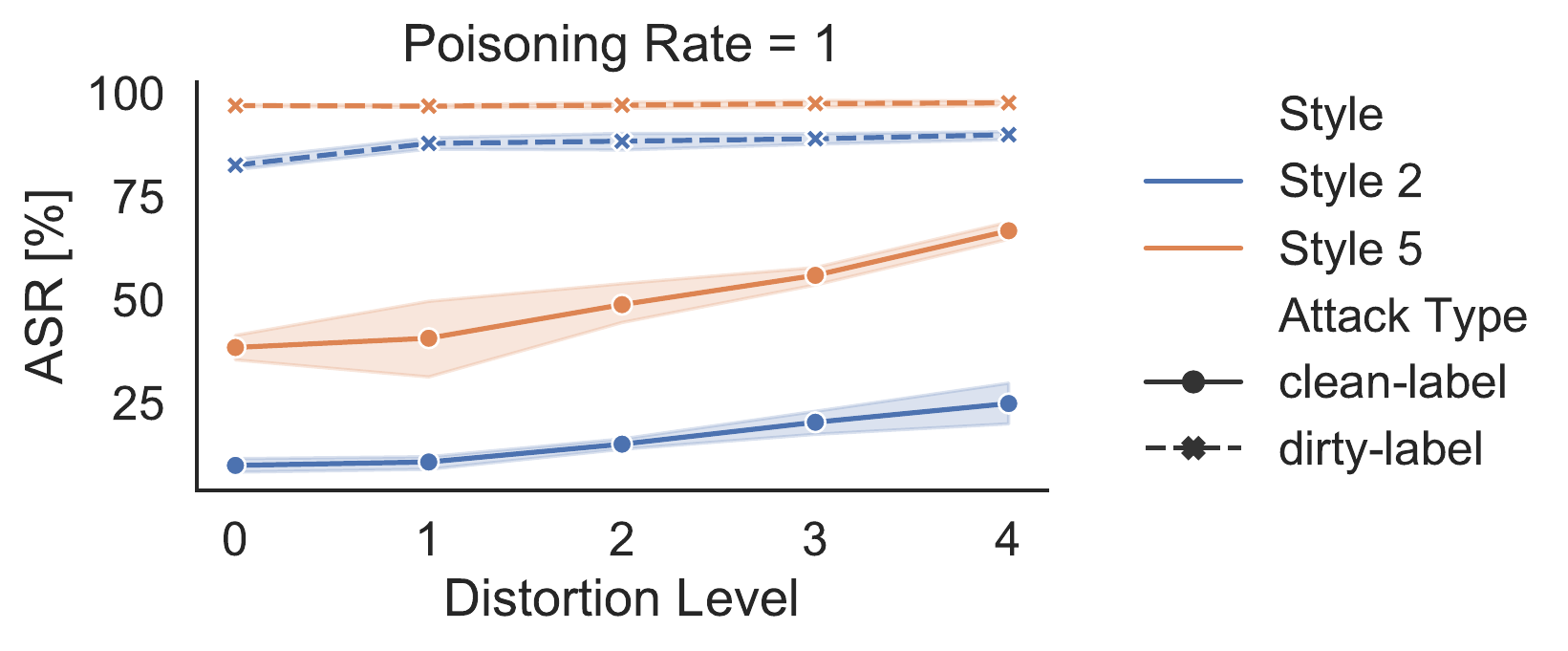}
    \caption{ASR for different distortion levels for styles 2 and 5.}
    \label{fig:distortions-level}
\end{figure}

To understand the effect of each style's parameter on ASR, we performed an ablation study for our most effective styles (styles 2 and 5) on GSC. In this study, we varied the chorus's amplitude for style 2 ($[2, 4, 6, 8, 10]$) and the gain level for style 5 ($[4, 8, 12, 16, 20]$). \Cref{fig:distortions-level} shows that larger distortion increases ASR. The effect is more visible when the attack performs poorly, i.e., for the clean-label attack.
Future investigations should analyze the trade-off between ASR and audio quality (or stealthiness).

\subsection{Evasion}
Cao et al.~\cite{stylefool} showed that -- in CV -- stylistic transformations can lead to evasion attacks. Following such intuition, we tested our malicious samples on unpoisoned models and analyzed the evasion rate (misclassification). Style 0 and style 3 are the most effective attacks (70\% evasion, on average); however, the user study showed that these styles degrade the samples' quality.
The other triggers show better evasion performance, e.g., style 5 with an evasion rate equal to 42.8\% (Large-CNN), 70.2\% (Small-CNN), and 50.7\% (LSTM).

\section{Generalization on Different Datasets}
We now demonstrate that the attack generalizes in different datasets. We use the large-CNN for styles 2 and 5 (the most effective ones), with a poisoning rate equal to 1\% for the dirty-label scenario. Other settings are the same as in Section~\ref{ssec:sound-setup}.
\par
\textbf{TIMIT}~\cite{timit}.
We use the dataset's core test set, consisting of 240 audio clips from 24 speakers (speaker classification).
Each sample is clipped to 1 second to maintain consistency with the experimental settings adopted in this paper.
This task is challenging (F1-score 0.57), but the results show that ASR is, on average, 41.4\% (with 9.4 std) and 77.4\% (with 4.5 std) for styles 2 and 5, respectively.
\par
\textbf{Keras Speaker Recognition}\footnote{\url{https://keras.io/examples/audio/speaker_recognition_using_cnn}} is a speaker classification task of 5 users. The task is easier than the previous one (F1-score 0.99). The results show that ASR is, on average, 89.9\% (with 5.3 std) and 99.5\% (with 0.6 std) for style 2 and style 5, respectively.

%% file: Sections/08-Conclusions.tex
\section{Conclusions and Future Work}
\label{sec:conclusions}

This work contribution is twofold: a formal description of stylistic backdoors and JingleBack, the first stylistic backdoor in the audio domain.
We demonstrated the potential of our attack through an extensive evaluation considering 444 models, reaching up to 96\% of the attack success rate.
Future investigations are required to better understand the stylistic backdoors in the audio domain, for example, by considering the impact of the target class or, more in general, how to find an optimal style effective in both clean and dirty-label settings.
Finally, further investigation is needed regarding the stylistic backdoor's behavior against state-of-the-art countermeasures.